\documentclass[aps,prb,onecolumn,superscriptaddress]{revtex4-2}
\usepackage{graphicx} 
\usepackage{dcolumn}
\usepackage{bm}
\usepackage{physics2}
\usepackage{amsmath}
\usephysicsmodule{ab}   
\usephysicsmodule{ab.braket} 
\usepackage{fixdif, derivative} 
\usepackage{booktabs}   
\usepackage{amssymb}    
\newcommand{\Yes}{\checkmark}
\newcommand{\No}{\(\times\)}

\usepackage{color}

\begin{document}

\title{Kapitza–Dirac interference of Higgs waves in superconductors}
\author{Daemo Kang}
\affiliation{Department of Applied Physics, The University of Tokyo, Hongo, Tokyo 113-8656, Japan}
\affiliation{Nordita, Stockholm University and KTH Royal Institute of Technology, Hannes Alfvéns väg 12, SE-106 91 Stockholm, Sweden}
\affiliation{Department of Physics and Institute for Materials Science, University of Connecticut, Storrs, Connecticut 06269, USA}

\author{Tien-Tien Yeh}
\affiliation{Nordita, Stockholm University and KTH Royal Institute of Technology, Hannes Alfvéns väg 12, SE-106 91 Stockholm, Sweden}
\affiliation{Department of Physics and Institute for Materials Science, University of Connecticut, Storrs, Connecticut 06269, USA}

\author{Takahiro Morimoto}
\affiliation{Department of Applied Physics, The University of Tokyo, Hongo, Tokyo 113-8656, Japan}

\author{Alexander V. Balatsky}
\affiliation{Nordita, Stockholm University and KTH Royal Institute of Technology, Hannes Alfvéns väg 12, SE-106 91 Stockholm, Sweden}
\affiliation{Department of Physics and Institute for Materials Science, University of Connecticut, Storrs, Connecticut 06269, USA}

\begin{abstract}
    We present a novel framework for controlling Higgs mode and vortex dynamics in superconductors using structured light. We propose a phenomenon analog of the {\em Kapitza-Dirac effect} in superconductors, where Higgs waves scatter off light-induced vortex lattices, generating interference patterns akin to matter-wave diffraction. We also find that the vortices enable the linear coupling of Higgs mode to the electromagnetic field.
    This interplay between light-engineered Higgs excitations and emergent vortex textures opens a pathway to probe nonequilibrium superconductivity with unprecedented spatial and temporal resolution. Our results bridge quantum optics and condensed matter physics, offering new examples of quantum printing where one uses structured light to manipulate the collective modes in correlated quantum fluids.
\end{abstract}

\maketitle

\section{Introduction}

The manipulation of quantum states with intense electromagnetic fields has emerged as a frontier in condensed matter physics~\cite{oka_floquet_2019,schlawin_cavity_2022}. Recent advances in this field include demonstration of light-induced superconductivity~\cite{mitrano_possible_2016}, Floquet-engineering topological phase ~\cite{oka_photovoltaic_2009}, and cavity-mediated superconductivity~\cite{schlawin_cavity-mediated_2019}. Quantum printing, the concept of transferring structured light features, further highlights the potential for manipulation of quantum matter ~\cite{aeppli_quantum_2025}. These examples illustrate how light can act not only as a probe but also as a powerful tool for control and creation of states of quantum matter. 

Beyond direct optical excitation, the Kapitza–Dirac (KD) effect, proposed in 1933~\cite{kapitza_reflection_1933}, describes the diffraction of a matter wave of electrons from a standing light wave acting as a periodic potential. The effect is reciprocal of the light(X-ray) scattering of the matter grid ( atomic crystal) in the classical Bragg scattering.  Initially conceived as a theoretical possibility, the effect was realized decades later in ultracold atoms~\cite{gould_diffraction_1986} and subsequently in free electrons~\cite{freimund_observation_2001}, firmly establishing its quantum mechanical foundations. Inspired by this phenomenon, we propose an analogous process in quantum matter- the superconducting film where Higgs waves, coherently excited by light, scatter from a light-induced vortex lattice. This scattering produces interference patterns that directly encode the spatial coherence of the Higgs mode, paralleling KD diffraction but with emergent quantum fluid excitations simultaneously serving as both the matter wave and the diffraction grating.

Additionally, light with spatial structure—often referred to as structured light~\cite{forbes_structured_2021,rubinsztein-dunlop_roadmap_2016}—further broadens the possibilities for controlling quantum materials by imprinting features onto the matter field. Such structured illumination has been proposed as a means to induce spiral spin-waves~\cite{fujita_encoding_2017}, topological defects in magnets~\cite{fujita_ultrafast_2017}, superconductors~\cite{yokoyama_creation_2020, yeh_quantum_2025,yeh_quantum_2025-1,yeh_light_2025}, control spin textures in spin-orbit coupled electrons~\cite{yamamoto_optical-vortex-pulse_2025}, lead magnetization in quantum Hall fluids~\cite{cardoso_orbital_2025}, and excite spiral Higgs mode~\cite{mizushima_imprinting_2023,kang_quantum_2025}. However, the imprinting effect of matter-wave interference elucidated by KD effect is still absent.  

Here, we focus on the Higgs modes and vortices in superconducting fluid and the transfer (imprinting) of spin and orbital momentum of light onto superconducting fluid. 
In superconductors, the Higgs mode corresponds to oscillations in the amplitude of the superconducting order parameter, analogous to the Higgs boson in particle physics~\cite{shimano_higgs_2020}. This collective excitation is gapped at twice the superconducting gap energy (\(2\Delta\)) and often appears as the lowest collective excitation mode in superconductors, which provides a unique probe of superconducting properties~\cite{schwarz_classification_2020}, and has been experimentally accessed through nonlinear optical processes such as third-harmonic generation~\cite{matsunaga_light-induced_2014}.  

In this work, we demonstrate that structured light can generate vortices in superconductors by solving the time-dependent Ginzburg–Landau equation with a second-order time derivative, a framework commonly employed to describe the dynamics of the Higgs mode~\cite{tsuji_theory_2015, tsuji_higgs_2024,mizushima_imprinting_2023,kang_quantum_2025}. By imprinting the spatial structure of light onto the superconducting order parameter, we show that the inclusion of inertial dynamics enables the coexistence of Higgs waves and vortices, and that Higgs excitations can serve as a sensitive probe of vortex dynamics through a novel linear response of the Higgs mode induced by light-induced vortices. Building on this, we establish an analogy to the Kapitsa–Dirac effect, where Higgs waves scatter from a light-induced vortex lattice and form interference patterns that directly reveal spatial coherence, Table \ref{tab:prior-work}. In our proposal we find the interplay interesting physical effects: i)   the light induced vortices - the "light induced" grid that scatters matter waves, ii) the generation of Higgs waves - the matter waves to scatter off light and iii) the scattering of the Higgs wave off the vortex core- the topological defect that enables the {\it linear} order (proportional to ${\bf A}(t)$) linear in frequency response of Higgs wave all come together in our proposal.

Together, these results expand the scope of quantum printing and provide a framework for manipulating and imaging nonequilibrium superconducting states with structured light, opening new avenues for exploring symmetry, coherence, and dynamics in quantum materials far from equilibrium.

\begin{figure}[t]
\includegraphics[width=\columnwidth]{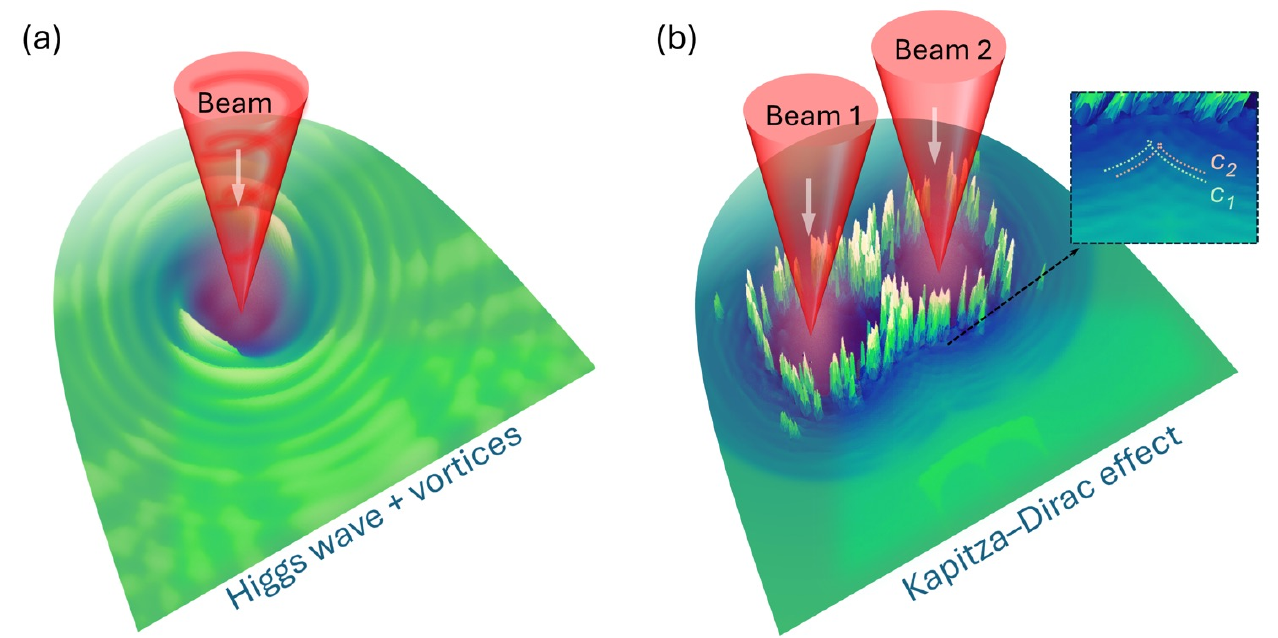}
\caption{\label{fig1} Schematic figures of Higgs mode and structured light-induced interaction. (a) Chiral Higgs wave propagation induced by circularly polarized light (\(s = 1\)) (b) Kapitza-Dirac effect, and Higgs mode interference diffracted by light-induced vortex pairs. The inset shows a magnified view of the Higgs-wave interference pattern, highlighting the regions marked by curves $c_1$ and $c_2$, corresponding to constructive and destructive fringes, respectively.}
\end{figure}

\begin{table}[t]
  \centering
  \caption{Overview of prior studies categorized by type of time-dependent Ginzburg–Landau (TDGL) formulation and associated physical phenomena. A checkmark indicates that the corresponding reference explicitly addressed the listed item.}
  \label{tab:prior-work}
  \setlength{\tabcolsep}{10pt}
  \renewcommand{\arraystretch}{1.2}
  \begin{tabular}{@{}lcc@{}}
    \toprule
    & Higgs mode & Vortices \\
    \midrule
    TDGL (type-I TDGL)  & \No  & \Yes\ \cite{yeh_quantum_2025,yeh_quantum_2025-1} \\
    TDGL (type-II TDGL)  & \Yes\ \cite{tsuji_theory_2015}      & \Yes   \\
    \bottomrule
  \end{tabular}
\end{table}

\section{Model}
In this study, we analyze the dynamics of superconductors using time-dependent Ginzburg-Landau (TDGL) equation. The TDGL equation phenomenologically describes the dynamics and spatial structure of the superconducting order parameter \(\Psi(\bm{r}) \). Here, we derive the TDGL equation from the Lagrangian of a \(s\)-wave superconductor:

\begin{equation}\label{eq:Lagrangian}
    \mathcal{L} = -\left[a|\Psi|^2+\frac{b}{2}|\Psi|^4+\frac{1}{2m^*}\left|\left(-i\hbar\nabla-e^* \bm{A}\right)\Psi\right|^2\right]  +c\left|i\hbar\frac{\partial \Psi}{\partial t}\right|^2 +d\Psi^{\ast}i\hbar\frac{\partial \Psi}{\partial t}
\end{equation}
where \(a\), \(b\), \(c\), and \(d\) are the phenomenological parameters that characterize the superconducting state, and \(m^*\)and \(e^*\) are the effective mass and charge of the Cooper pairs, respectively. Near the critical temperature \( T_c \), the coefficient \( a \) takes the form \( a = a_0 (T - T_c) \), where \( a_0 > 0 \). 
Under equilibrium conditions, the order parameter attains the value \( \Psi_0 = \sqrt{-a/b} \) with \( b > 0 \), yielding a superfluid density of \( n_s = |\Psi_0|^2 = -a/b \). 
The parameters \( c \) and \( d \) appear in the time-dependent part of the Lagrangian and control the dynamics of the order parameter. In superconductors, particle-hole symmetry at low energies allows the \(d\)-term to be zero ~\cite{tsuji_theory_2015, tsuji_higgs_2024}. Applying the variational principle, we obtain the TDGL equation. To account for the damping of Higgs mode oscillations, we phenomenologically introduce an additional term and consider the following modified equation.
\begin{equation}
    c\hbar^2\pdv[2]{\Psi}{t} +  \Gamma\pdv{\Psi}{t} = -\ab[a\Psi +  b\Psi|{\Psi}|^2 +\frac{1}{2m^*}(-i\hbar\nabla-e^{*}\bm{A})^2\Psi ] 
\end{equation}

To arrive at TDGL equation in dimensionless form, Eq. \ref{eq:nondimTDGL}, we use rescaled parameters: the order parameter is normalized as \( \psi = \Psi / \Psi_0 \), time is rescaled as \( t' =   \Omega_{\mathrm{Higgs}}\, t \), $\tau=1/2$ where the Higgs mode frequency is defined by \( \hbar \Omega_{\mathrm{Higgs}} = \sqrt{-a / (2c)} \) = \( \Delta_0 \) (Superconducting gap). The spatial coordinate is scaled by the coherence length \( \xi = \hbar / \sqrt{4 m_e |a|} \), giving \( \bm{r}' = \bm{r} / \xi \), and the vector potential is rescaled as \( \bm{A}' = \bm{A} / A_0 \), with \( A_0 = \hbar / (2e \xi) \),.  The initial state is taken as a uniform superconducting order parameter in the ground state $\psi=1$ (corresponding to $a=-1, b=1$).
In this numerical calculation, we employ the open (Dirichlet) boundary condition.

\begin{equation}\label{eq:nondimTDGL}
    \tau\frac{\partial^2 \psi}{\partial t'^2} + \gamma\pdv{\psi}{t'} =  \psi - \psi|\psi|^2 + (\nabla'- i\bm{A}')^2 \psi
\end{equation}

This model differs from the first-order time-dependent Ginzburg–Landau (g-TDGL~\cite{kramer_theory_1978, watts-tobin_nonequilibrium_1981,bishop-van_horn_pytdgl_2023}) framework employed in Refs.~\cite{yeh_quantum_2025, yeh_quantum_2025-1}. The first-order TDGL describes overdamped, diffusive dynamics analogous to a relaxation equation of motion, where excitations are overdamped and no oscillations are possible.  The inclusion of a second-order time derivative transforms the equation into a wave-like form and allows for the oscillatory solutions. The respective characteristic timescales of the system are occur at different time scales: the diffusive dynamics typically occur on the microsecond scale, whereas inertial (wave-like) dynamics appear on the picosecond scale under THz laser driving. This inertial extension enables the propagation of collective modes such as the Higgs oscillation and allows us to analyze the real-time dynamics of light-driven vortices.

As the light source, we employ structured light, specifically a Laguerre–Gaussian (LG) beam. This beam mode is characterized by three parameters: $s, l$, and $p$, which correspond to the spin angular momentum (related to the polarization), the orbital angular momentum, and the radial index, respectively~\cite{allen_orbital_1992}. The vector potential of the LG beam in its general form is given as follows:
\begin{equation}
    \bm{A}^{(i)}_{s,l,p}(\bm{r
    },t) = \textrm{Re}\ab[A_0\bm{n}_su_{l,p}(r,\phi)\exp(-i\Omega t)f(t)]    
\end{equation}

\begin{equation}       
     u_{l,p}(r,\phi) =  \sqrt{\frac{2p!}{\pi(p+|l|)!}}\left(\frac{\sqrt{2}r} {w_0}\right)^{|l|} e^{-\frac{r^2}{w_0^2}}e^{il\phi}L^{|l|}_p\left(\frac{2r^2}{w_0^2}\right).
\end{equation}
Here, we employ the continuous beam (function $f(t)=1$ ). Parameters $\Omega$, $\phi$, and $w_0$ in represent the frequency of light, azimuthal angle, and spot size, respectively. The function $L^{|l|}_p$ is Laguerre polynomial. $\bm{n_s}$ represents the polarization of light. In this study, the linearly polarized light is $ \bm{n_s}=\bm{e_x}$ $(s=0)$ and circularly polarized light is $\bm{e_s}=(\bm{e_x}\pm\bm{e_y})/\sqrt{2}$ $(s=\pm1)$.

In this study, we consider illuminating the system with multiple beams. The total vector potential of the matter field is then given by
\begin{equation}
    \bm{A}(\bm{r},t) = \sum_i \bm{A}^{(i)}_{(s, l, p)}(\bm{r}-\bm{r_{0,i}},t)=\bm{A}^{(1)}_{(s, l, p)}(\bm{r}- (-d\times w_0) \bm{e_x},t)+\bm{A}^{(2)}_{(s, l, p)}(\bm{r}- (+d\times w_0)\bm{e_x},t) ,
\end{equation}
where each \( \bm{A}^{(i)}_{(s, l, p)} \) represents an individual Laguerre–Gaussian beam characterized by the indices \(s\), \(l\), and \(p\), the distance of each two laser beam is defined by $ w_0 \times d$ in the following context, and $\bm{r_{0,i}}$ indicates the center of the $i^{th}$ beam. 

For the creation of Abrikosov vortex lattice, we introduced uniform and static magnetic field $\bm{B}=\nabla\times\bm{A}$ with $|\bm{A}|=0.02$. After forming a stable vortex lattice,
at $\Omega_{Higgs}t/2\pi=20$ and having a stable vortex lattice, we illuminate the system with structured light of amplitude $|\bm{A}|=0.2$ to investigate the interaction between vortex lattice and Higgs mode.

To characterize the collective excitation, we perform a spatiotemporal Fourier transform of the complex order parameter $\psi(x,y,t)$: $\psi(q_x,q_y,\omega)=FFT_{x,y,z}[\psi(x,y,t)]$. The Higgs and Nambu-Goldstone (NG) modes typically appear as a parabolic dispersion near 2$\omega$ and a linear dispersion near $\omega$=0, respectively. In this model, we don't include the charge pumping of the phase mode, so that the phase mode remains in the plot even though it goes into the plasma frequency by the Anderson-Higgs mechanism.

Considering niobium nitride (NbN) as an example of \(s\)-wave superconductor, superconducting gap of would be \(2\Delta_0 ~\approx O( 2~\mathrm{meV})\) and a coherence length of \(\xi \approx O( 4~\mathrm{nm})\), so that a thin-film sample dimension \(100\xi\) corresponds to \(O(400~\mathrm{nm})\)~\cite{matsunaga_light-induced_2014}. In this case, the Higgs resonant frequency of light is estimated to be \(\Omega \approx O(1~\mathrm{THz})\), and a beam fluencey corresponding to \(|A|/A_0 = 1\) gives an electric field amplitude of approximately \(O(1~\mathrm{kV/cm})\).
The dynamical results with movies are provided in the supplementary materials\cite{supp_DK}.

\section{Results}
\begin{figure}[t]
\includegraphics[width=0.9\textwidth]{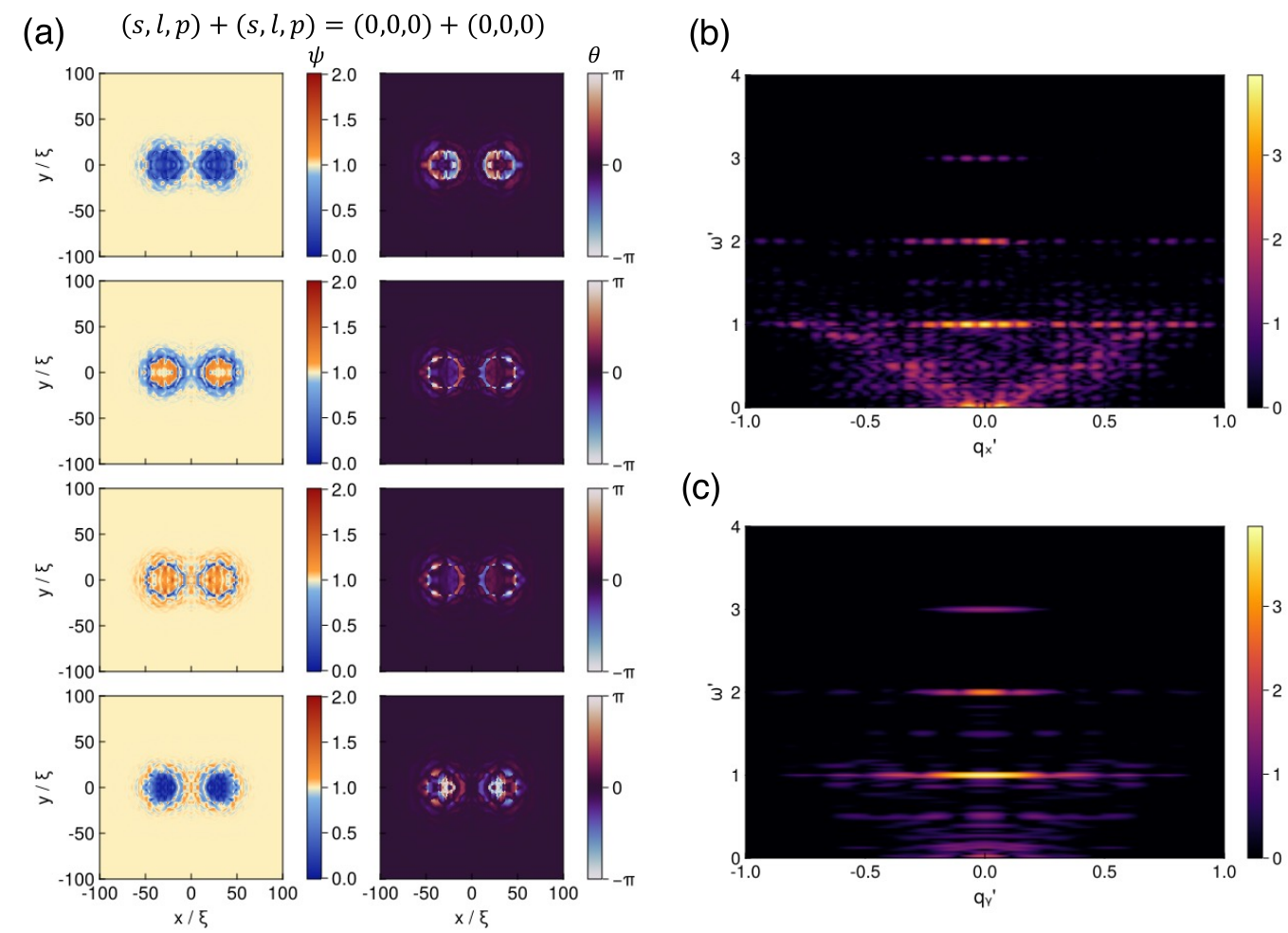}
\caption{\label{fig2} Kapitza-Dirac effect of Higgs mode (matter wave) diffracted by structured light-induced vortex lattice (optical lattice).  (a) Real-space interference patterns of the Higgs-mode amplitude and phase at time frames $\Omega_{Higgs}t/2\pi=24.55, 24.75, 24.85, 25$. 
(b, c) Corresponding spectrum (log-scale) of $\psi(q_x,0,\omega)$, $\psi(0,q_y,\omega)$ ( $\omega'=\omega/\Omega_{Higgs}$, $q'=q\xi/2\pi$ ). The damping parameter is $\gamma=0.1$, and the amplitude of vector potential is $|\bm{A'}|=2.0$.
}
\end{figure}
To set up the KD in superconductors, we use {\em two} beams (distance $d=2$ and beam size $w_0=20$). While the effect is reciprocal in both beams one can view the first beam as the source of vortices (= grid for the incoming Higgs waves). The second beam serves as the source of Higgs waves resonantly excited by light. By illuminating the superconductor with two intense laser beams, both the interference pattern and the vortex lattice become visible simultaneously, as shown in Figure 2(a). 
As a result, complex spatial structures emerge, representing the diffraction of Higgs waves by the structured-light–induced vortex lattice.
Figures 2(b) and 2(c) show that the coexistence of vortex pairs and Higgs-mode interference gives rise to both interference stripes in the (\(q_x–\omega\)) and (\(q_y-\omega\)) dispersion relation and a linear coupling originating from vortex creation. Especially, the interference stripes in $q_y$ suggest the diffraction from the vortex lattice.

The presence of vortices enables the linear coupling of the Higgs mode to the electromagnetic field. In most cases, the Higgs mode does not exhibit a linear coupling; however, the introduction of vortices generates phase singularities that prevent complete gauge removal of the phase mode, taking unitary gauge $ eA_{\mu}\to eA_\mu - \partial_{\mu}\theta$, and allowing a new linear response in the kinetic term as NG mode and Higgs mode coupling. Writing the order parameter as $\psi = |\psi|e^{i\theta_s}$, in TDGL equation one obtains
\begin{align}
i\mathbf{A}'\nabla'\psi 
&= i\mathbf{A}'e^{i\theta_s}\nabla'|\psi| - \mathbf{A}'\psi\nabla'\theta_s .
\end{align}
This second term suggests the coupling of the amplitude (Higgs) mode and the phase (NG) mode. Noting that the linear dispersion of the NG mode and other peaks in Figure 2 (b)(c) cannot be observed due to the Anderson-Higgs mechanism.
To observe the Kapitza-Dirac effect, one possible approach is to observe the interference stripes of $q_y$ in the linear response.

\begin{figure*}[t]
\includegraphics[width=0.9\textwidth]{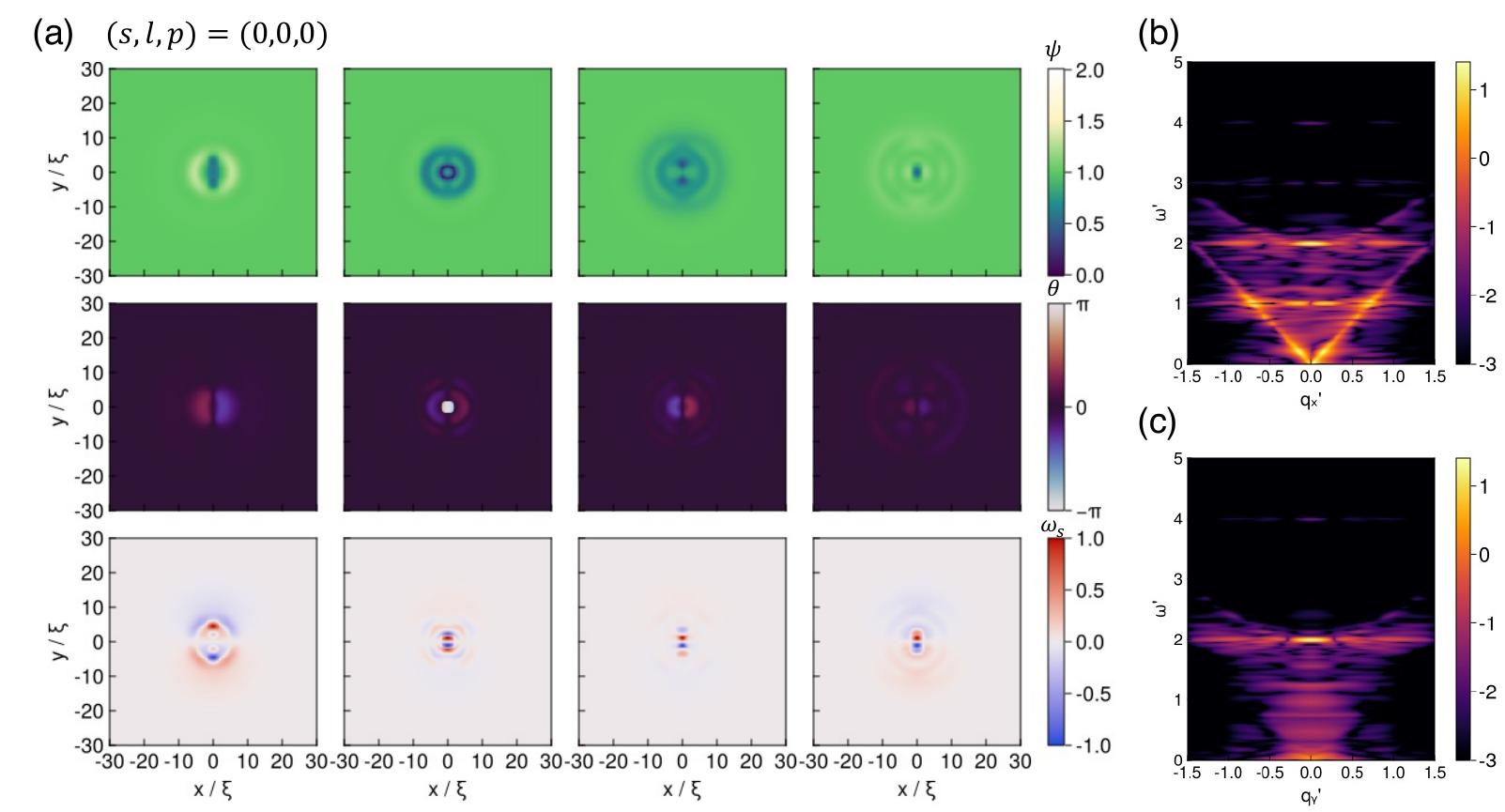}
\caption{\label{fig3} Snapshots showing vortex creation and dispersion relation suggesting Higgs and Nambu–Goldstone (NG) modes coupling.  (a) Time evolution of amplitude, phase, and vorticity at time frames \(\Omega_{Higgs}t/2\pi=2.01,2.31,2.61\), and \(2.91\). (b,c) Dispersion relation in the $\psi(q_x,0,\omega)$, $\psi(0,q_y,\omega)$ plane (log-scale). Parameters: $\gamma=0.05,|\bm{A}'|=1.0,w_0=10$
}
\end{figure*}

\begin{figure*}[t!]
\includegraphics[width=0.9\textwidth]{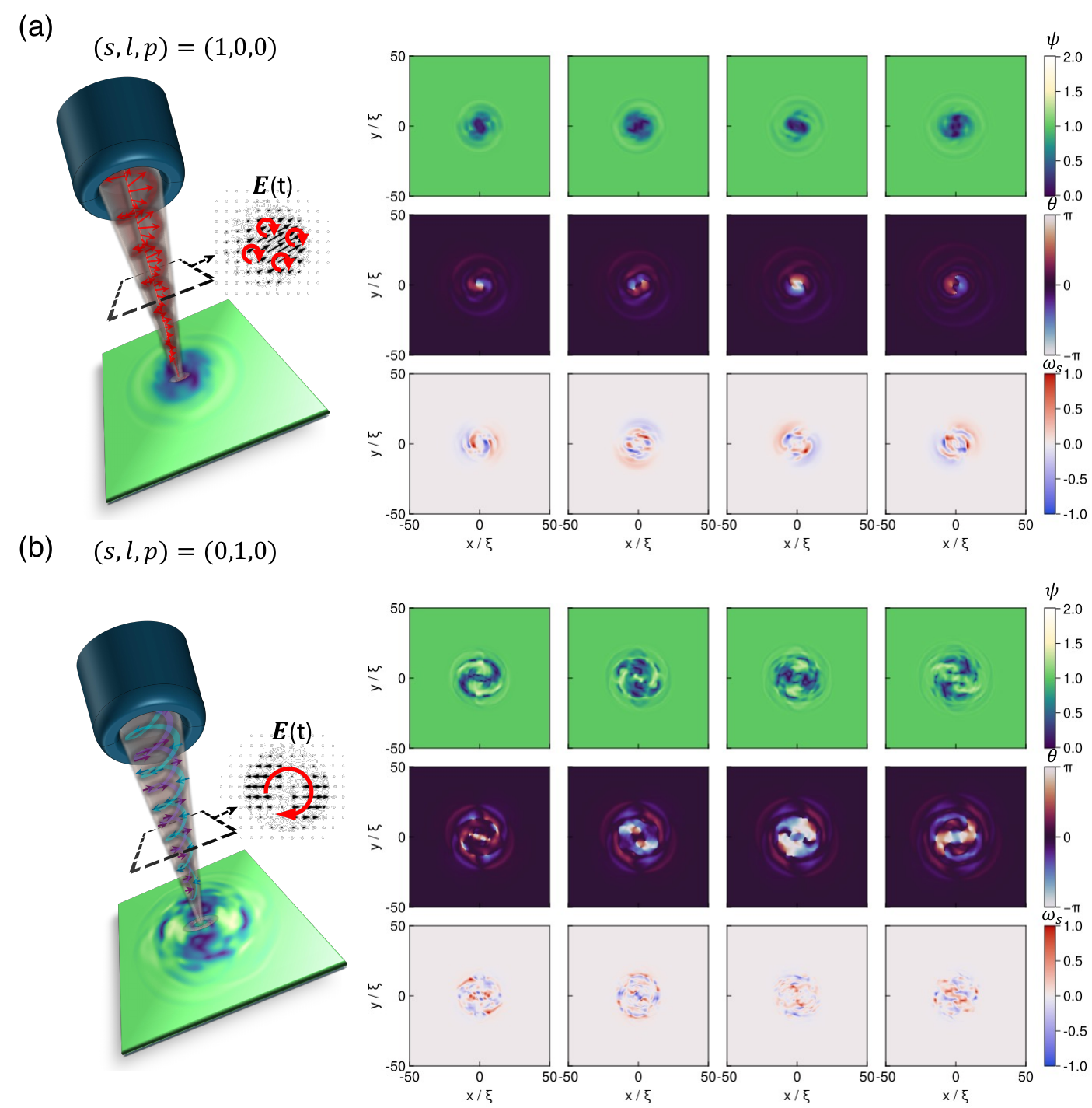}
\caption{\label{fig4} Snapshots showing the spiral dynamics of vortices and Higgs mode driven by light with the angular momentum. 
(a, b) Time evolution of the amplitude, phase, and vorticity of the order parameter under illumination with a circularly polarized Gaussian beam \((s, l, p) = (1, 0, 0)\) and Laguerre–Gaussian (LG) vortex beam \((s, l, p) = (0, 1, 0)\) at time frames \(\Omega_{Higgs}t/2\pi=2.01,2.31,2.61\), and \(2.91\). Parameters:$\gamma=0.05,|\bm{A}'|=2.5,w_0=10$
 }
\end{figure*}

To illustrate the coupling between the Higgs and NG modes in the presence of vortices, we demonstrate the creation of vortices by structured light in Figures 3 and 4. Figure 3 (a) shows the time evolution of vortex-antivortex pairs generated under Gaussian beam illumination, where the supercurrent $J_s=\mathrm{Im}[\psi^*(\nabla-i\bm{A})\psi]$ and vorticity $\omega_s = \nabla\times\bm{J_s}$.
The oscillating (AC) magnetic field produced by the structured light suppresses the order parameter amplitude, leading to the nucleation of vortex pairs when the field strength becomes sufficiently large. 
Remarkably, this vortex creation appears even in the absence of the relaxation term (\(\gamma=0\)), which is the dominant dynamical term in Ref~\cite{yeh_quantum_2025,yeh_quantum_2025-1}. This indicates that the mechanism of vortex generation is fundamentally driven by the magnetic-field dynamics rather than by dissipative relaxation. Furthermore, the inclusion of the inertia term allows the created vortex to induce the fluctuation of the amplitude mode as well, which is otherwise overdamped in the conventional g-TDGL model and enables the simulation of the Kapitza-Dirac effect. Figure 3(b)(c) show the dispersion (\(q_{x,y}–\omega\)) of amplitude of the order parameter. The parabolic curve in Fig.3(b) around $\omega'=2$ shows the dispersion of the Higgs mode, and as we discussed above, the linear coupling of the Higgs mode could become observable due to the existence of light-induced vortices.

With clear evidence of strong coupling between the Higgs dynamics and vortices, the spiral motion of vortices, leading to complex angular-momentum transfer, becomes an intriguing phenomenon to explore.
Figure 4 shows the spiral dynamics of vortices and Higgs mode driven by the light with the angular momentum. Fig.4.(a) presents the rotation of a vortex-antivortex pair, accompanied by a swirling pattern of the Higgs mode. The OAM of the LG beam can also induce the rotational Higgs mode, showed in Fig.4.(b); however, because the magnetic field of LG beam is more complex and not spatially continuous, unlike that of a circularly polarized Gaussian beam, the resulting dynamics exhibit alternating confinement and deconfinement of vortices.

\begin{figure*}[t!]
\includegraphics[width=0.9\textwidth]{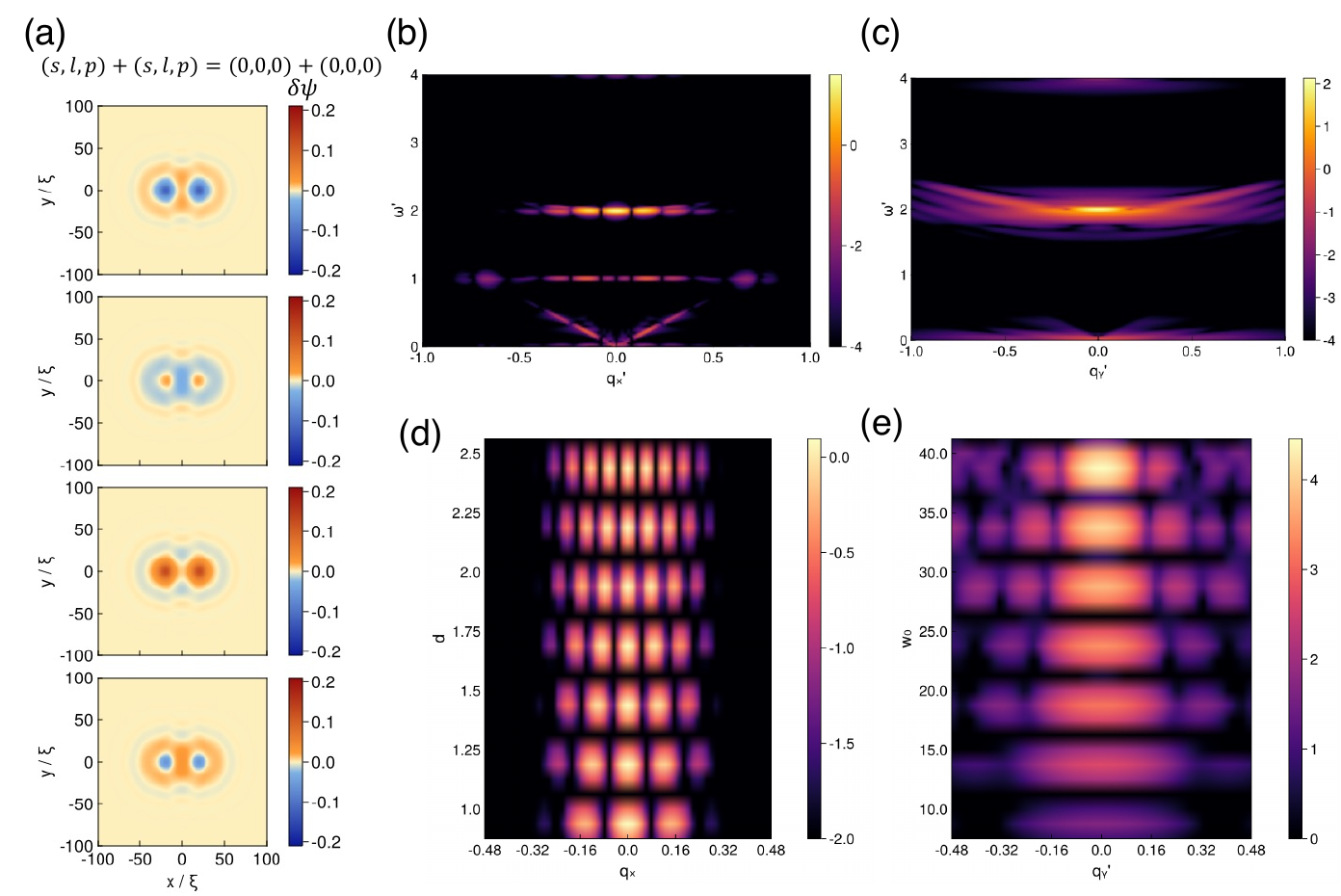}
\caption{\label{fig5} 
Snapshots showing the interference of the Higgs mode induced by two laser beams incident on the superconductor. 
(a) Real-space interference patterns of the Higgs-mode amplitude and phase at time frames $\Omega_{Higgs}t/2\pi=24.55, 24.75, 24.85, 25$. Parameters:$\gamma=0,|\bm{A}'|=0.1,w_0=10$
(b, c) Corresponding  $\psi(q_x,0,\omega)$, $\psi(0,q_y,\omega)$spectrum (log-scale). 
(d) Dependence of the Higgs-mode interference in $q_x$ at $\omega' = 2$ on the separation $d$ between the two laser beams ($w_0 = 10$).
(e) Dependence of the Kapitza--Dirac interference in $q_y$ (linear-coupling regime ($\omega'=1$) on the beam size $w_0$ ($|\bm{r}_{0,i}|=2\times20=40$ is fixed).
}
\end{figure*}

Figure 5 demonstrates the interference of the Higgs mode in superconductors when we introduce the weaker two laser beams ($|\bm{A}'|=0.1)$ and do not create vortices. Figure 5(a) illustrates the real-space interference pattern of the Higgs mode. Notably, the central region of the fluctuation occupies a finite area, and as the beam size increases, the interference fringes become less distinguishable. Even in the absence of the dissipation term, the Higgs propagation does not persist infinitely but decays due to the nonlinear term in the TDGL equation. Figure 5(b)(c) show the corresponding ($q_{x,y}-\omega$) spectra. In Figure 5(b), the $q_x$ cut suppresses specific wave number components, producing a stripe-like feature, whereas the $q_y$ cut does not exhibit such interference features. This is because the laser spots are arranged along the x-axis. In contrast, Figure 2(c) shows the interference pattern in the $q_y$ because of the diffraction of the vortex lattice along the y direction. This directional dependence of the Higgs wave interference provides evidence for the KD effect induced by the structured light-generated vortex lattice.
Figure 5(d)(e) supports the trend that the spacing of the interference stripes depends on the separation of the grid: as the distance increases, the suppressed wave number shifts to smaller values. Figure 5(d) shows the dependence of the Higgs mode interference on the beam distance, demonstrating the location of the canceled wave number is directly controlled by the beam separation. Figure 5 (e) shows the KD interference of $\omega'=1$ with the different beam sizes. As the beam size becomes wider, the resulting vortex lattice expands, causing the KD interference stripes to shift accordingly.

\begin{figure}[t!]
\includegraphics[width=\columnwidth]{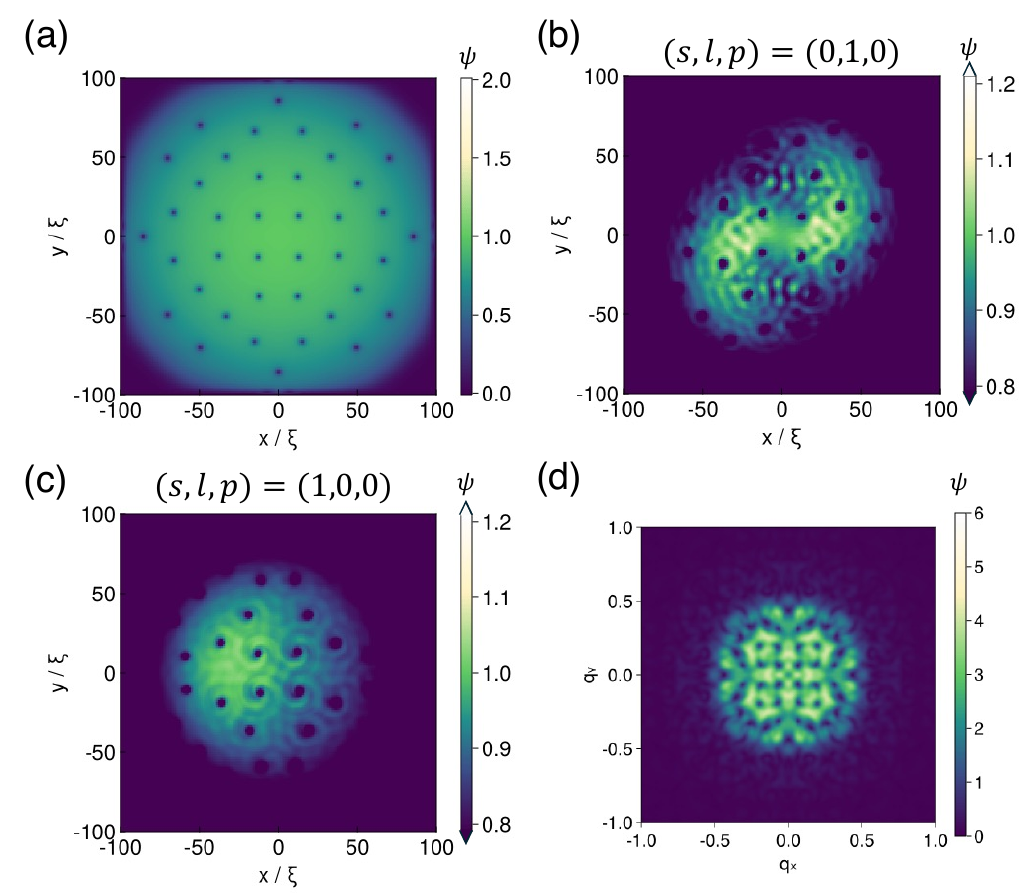}
\caption{\label{fi6} 
Vortex lattice and Higgs wave patterns induced by structured light. (a) Vortex lattice forms under a static uniform magnetic field. (b) and (c)  Higgs mode diffraction from the vortex lattice driven by (b) LG beam (\((s,l,p)=(0,1,0)\)) and (c) CPL (\((s,l,p)=(1,0,0)\)). (d) Real space Fourier transformation of $\psi(q_x,q_y)$ spectrum (log-scale).}
\end{figure}

Moreover, the vortex-enhanced Higgs-mode oscillation is not solely induced by optically generated vortices. Figure 6 illustrates the formation of a vortex lattice under a static magnetic field and its interaction with the Higgs mode. By applying a static and uniform magnetic field, the superconductor forms an Abrikosov (vortex) lattice, as shown in Fig. 6 (a). The number and spatial distribution of vortices depend on the field strength and the dissipation coefficient \(\gamma\), which determine the extent of the superconducting region. Figure 6 (b) shows the vortex lattice under illumination by an LG beam, leading to the excitation of a spiral Higgs wave that diffracts from the vortex array. Figure 6 (c) presents the case of circularly polarized light, where the spin angular momentum of light drives a chiral Higgs wave around each vortex core. Figure 6 (d) shows the reciprocal space of the vortex lattice, revealing the characteristic periodicity corresponding to the vortex spacing. This result demonstrates that the present framework provides a versatile platform to investigate the interplay between vortex dynamics and light fields in superconductors.

The stability of the Higgs mode can be discussed by considering the energy dispersion of the Higgs mode for finite momentum, $ \hbar\omega (\bm{q}) = \sqrt{(2\Delta_0)^2+\hbar^2v_F^2q^2/3}-i\pi^2\hbar v_Fq/24$~\cite{littlewood_amplitude_1982}, where $v_F$ is the Fermi velocity and amplitude of momentum $q=|q|$. The momentum of the Higgs mode obtained in this calculation satisfies $\hbar v_Fq/\Delta\lesssim1$, so this Higgs mode interference should be feasible, such as in s-wave superconductor NbN.~\cite{chockalingam_superconducting_2008}.  To observe this phenomenon, one approach is to use the spatially-resolved pump-probe technique and see the dynamics of the excited Higgs mode.

\section{Conclusion}

By merging concepts from quantum optics and superconductivity in quantum printing, we demonstrate that structured light can sculpt Higgs mode landscapes and induce Kapitza-Dirac scattering in a quantum fluid.
We discussed the emergent effects as the result of the light induced vortices and scattering of Higgs waves:  the light induced Higgs waves interfere depending on the structure of the light and the scattering of the Higgs wave of the vortex cores.  We also find the linear in the ${\bf A}$ scattering that can be see in spectroscopy. All of these results represent interesting phenomena that can be quantum printed in superconductors using structured light.  We also point that quantum printing paradigm extends to other collective modes (e.g., charge density waves), offering a universal tool for probing and manipulating quantum order, something we plan to explore in futuer work.

\section{Acknowledgment}

We are grateful to Y. Liu and H. Yerzhakov for useful discussions. 

This work was supported by U.S. Department of Energy, Office of Science, Office of Basic Energy Sciences under Award No. DE-SC-0025580 (A.V.B. and TTY – conceptualization, calculation, writing). D. Kang was supported by the JST SPRING, Grant Number JPMJSP2108 and Forefront Physics and Mathematics program to drive transformation (FoPM) and  by European Research Council under the European Union Seventh Framework ERC-2018-SYG 810451 HERO (conceptualization, calculation, writing, travel).   
T.M. was supported by 
JSPS KAKENHI Grant 23K25816, 23K17665, and 24H02231 (conceptualization, calculation).

\bibliographystyle{unsrt}
\bibliography{Kapitza-Dirac_effect}

\end{document}